\documentclass[aps,10pt,twocolumn,superscriptaddress,footinbib,flushbottom]{revtex4-1}%
\usepackage{amssymb,amsmath,amsfonts,bm,dcolumn}
\usepackage{graphicx,color}
\usepackage[squaren]{SIunits}
\usepackage[english]{babel}
\usepackage{tabularx}
\usepackage{tikz}

\renewcommand{\vec}[1]{\boldsymbol{\mathrm{#1}}}%
\newcommand{\uvec}[1]{\bm{\hat{\mathbf{#1}}}}
\newcommand{\abs}[1]{\lvert#1\rvert}%

\newcommand{\mean}[1]{\langle #1 \rangle}

\renewcommand{\O}[1]{\mathcal{O}\left(#1\right)}

\newcommand\id{\bm{\ensuremath{\mathbb{I}}}}

\newcommand{\dif}{d}

\newcommand{\integral}[4]{\int_{#1}^{#2} \! #3 \, \dif#4}%

\newcommand{\ii}{\mathrm{i}}
\renewcommand{\Re}{\fct{Re}}%

\newcommand{\DiracDelta}[1]{\operatorname{\delta}\left(#1\right)}
\newcommand{\fct}[1]{\operatorname{#1}}

\newcommand*\showwidth[1]{%
	\textcolor{blue}{\rule{\csname#1\endcsname}{1pt}}\newline
	\texttt{\textbackslash#1}: \expandafter\the\csname#1\endcsname
	\par
}

\newmuskip\pFqmuskip
\newcommand*\pFq[6][8]{%
	\begingroup 
	\pFqmuskip=#1mu\relax
	\mathchardef\normalcomma=\mathcode`,
	\mathcode`\,=\string"8000
	\begingroup\lccode`\~=`\,
	\lowercase{\endgroup\let~}\pFqcomma
	{}_{#2}F_{#3}{\left[\genfrac..{0pt}{}{#4}{#5};#6\right]}%
	\endgroup
}
\newcommand{\pFqcomma}{{\normalcomma}\mskip\pFqmuskip}

\begin{document}
	
	\title{Active Brownian motion with memory delay induced by a viscoelastic medium}
	
	\author{Alexander R. Sprenger}
    \email{sprenger@thphy.uni-duesseldorf.de}
	\affiliation{Institut f\"{u}r Theoretische Physik II: Weiche Materie, Heinrich-Heine-Universit\"{a}t D\"{u}sseldorf, D-40225 D\"{u}sseldorf, Germany}
	
	\author{Christian Bair}
	\affiliation{Institut f\"{u}r Theoretische Physik II: Weiche Materie, Heinrich-Heine-Universit\"{a}t D\"{u}sseldorf, D-40225 D\"{u}sseldorf, Germany}
	
	\author{Hartmut L\"{o}wen}
	\affiliation{Institut f\"{u}r Theoretische Physik II: Weiche Materie, Heinrich-Heine-Universit\"{a}t D\"{u}sseldorf, D-40225 D\"{u}sseldorf, Germany}

	\date{\today}

	\begin{abstract}
		By now active Brownian motion is a well-established model to describe the motion of mesoscopic self-propelled particles in a Newtonian fluid. 
		On the basis of the generalized Langevin equation, we present an analytic framework for active Brownian motion with memory delay assuming time-dependent friction kernels for both translational and orientational degrees of freedom to account for the time-delayed response of a viscoelastic medium. 
		Analytical results are obtained for the orientational correlation function, mean displacement and mean-square displacement which we evaluate in particular for a Maxwell fluid characterized by a kernel which decays exponentially in time.
		Further, we identify a memory induced delay between the effective self-propulsion force and the particle orientation which we quantify in terms of a special dynamical correlation function.
		In principle our predictions can be verified for an active colloidal particle in various viscoelastic environments such as a polymer solution.
	\end{abstract}

	\maketitle

	\section{Introduction}
	
	The physics of active matter is a booming research area exploring non-equilibrium phenomena of self-propelled particles \cite{BechingerDLLRVV2016,GompperEtAl2020}.
	Apart from viscous damping in a fluid medium, fluctuations become important if the particle size is on the mesoscopic colloidal scale.
	A by now well-established model to describe the persistent random dynamics of a single self-propelled particle is so-called "active Brownian motion" \cite{RomanczukBELSG2012,BechingerDLLRVV2016,CallegariG2019,Loewen2020,GompperEtAl2020,HechtUL2021,LiebchenM2021}. 
	Here the translational coordinate of the particle is coupled to its self-propulsion direction which is the orientational degree of freedom establishing basically a persistent random walk.
	Active Brownian motion assumes an instantaneous friction which is a well-justified assumption for a Newtonian background fluid, or in other terms, there is no memory effect of the medium. 
	However, in many situations, self-propelled or swimming particles are exposed to environments different to a Newtonian fluid \cite{QiuEtAl2014, FuPW2007, ShenA2011, JakobsenNMOCD2011, GagnonKA2014, LiuPB2011, Schwarz-LinekVCCMMP2012, ZhuLB2013, RileyL2014, ElfringL2015, DattNHE2017,LiLA2021}.
	Important examples for non-Newtonian backgrounds offered to self-propelled particles are polymer solutions \cite{MartinezSRWMP2014, PattesonGGA2015, ZoettlY2019, QiWGW2020, LiuSMW2021}, crystalline \cite{KriegerSP2014, vdMeerFD2016, BrownVDVSLLP2016} or liquid crystalline \cite{ZhouSLA2014, Lavrentovich2016, MushenheimTTWA2014, HernandezNTIMS2015, KriegerDP2015, KriegerSP2015, TrivediMASW2015, TonerLW2016, FerreiroCTLW2018} environments or even biologically relevant backgrounds such as mucus \cite{SleighBL1988, SuarezP2005}, dense tissues \cite{JosenhansS2002} or soil \cite{Wallace1968}.
	
	In this paper, we use an extended model for active Brownian motion in a viscoelastic medium. 
	In doing so we assume memory effects of the solvent via a friction kernel for both translational and orientational degrees of freedom besides fluctuations. 
	In fact, there are different models for active Brownian motion with memory effects induced by the surrounding medium \cite{HuWA2017, PeruaniM2007, DebnathGLML2016, GhoshLMM2015, NarinderBGS2018, SevillaRGS2019, GomezSBB2016, LozanoGSB2018, LozanoGSB2019, SaadN2019, NarinderGSB2019,MitterwallnerLN2020,MuhsinSS2021} and for passive Brownian motion in a viscoelastic medium \cite{IndeiSCP2010,Brader2010,GrimmJF2011,RaikherRP2013,BernerMGSKB2018,DoerriesLK2021}. 
	Here we include activity explicitly.
	In contrast to Ref.~\cite{SevillaRGS2019} where an active Ornstein-Uhlenbeck approach is chosen and to Ref.~\cite{MitterwallnerLN2020} where negative friction is used to achieve activity, we chose our model to recover the  established active Brownian motion case for a Newtonian medium as a  clear reference state.
	In particular, the model used here is a special case of that recently proposed by Narinder and coworkers \cite{NarinderBGS2018} which contains an additional term of translation-rotation coupling between the swim force and the swim torque. 
	We consider here the special case of decoupled effective swim force and swim torque with the benefit that we can solve the stochastic Langevin equations analytically.
	We  evaluate the solution in particular for a Maxwell fluid which is characterized  by a kernel that decays exponentially in time and obtain analytical results for the mean displacement, the mean-square displacements and the orientational correlation function. 
	Further we define a memory delay function which measures the memory induced delay between the effective driving force and particle orientation.  
	In principle our predictions can be verified for an active colloidal particle in various viscoelastic environments such as a polymer solution.

	The paper is organized as follows: The model is introduced and discussed in chapter II. 
	In chapter III general results are listed. 
	The solution is evaluated further for a generalized Maxwell (or Jeffrey) kernel with a memory exponentially decaying in time in chapter IV. 
	Finally we conclude in chapter V.

	\section{The model} \label{sec:Model}

	In our model, we consider a colloidal self-propelled particle in two spatial dimensions moving at a constant speed $v_0$ along its orientation $\uvec{n}(t)$ through a fluid with memory properties.
	We describe the state of the particle by its position $\vec{r}(t)$ and its angle of orientation $\phi(t)$ which denotes the angle between the orientation vector $\uvec{n}(t) = (\cos\phi, \sin \phi)$ and the positive $x$ axis, at the corresponding time $t$.
	The time-delayed response of the fluid is incorporated in the model in terms of a translational memory kernel $\Gamma_T (t)$ and a rotational memory kernel $\Gamma_R (t)$ which directly couple to the translation and rotation of the particle, respectively. 
	To further model circle swimming, we also include an effective swim torque which acts on the particle and leads to a circling frequency $\omega_0$. %
	On the basis of the generalized Langevin equation, the overdamped Brownian dynamics of the particle is described by the following coupled non-Markovian Langevin equations
	\begin{subequations} \label{eq:langevin}
		\begin{gather}
			\integral{-\infty}{t}{ \Gamma_T(t-t') \big( \dot{\vec{r}}(t') -  v_0 \uvec{n}(t') \big) }{ t' }  = \vec{\xi}(t), \label{eq:langevin_r} \\
			\integral{-\infty}{t}{ \Gamma_R(t-t') \big(\dot{\phi}(t') - \omega_0 \big) }{ t' }  = \eta (t), 	\label{eq:langevin_phi}
		\end{gather}
	\end{subequations}
	where $\vec{\xi}(t)$, $\eta(t)$ denote zero-mean Gaussian colored noise
	\begin{subequations}
		\begin{gather}
			\mean{\vec{\xi}(t)} = 0, \qquad \mean{\vec{\xi}(t) \otimes \vec{\xi}(t')}  = \id \, k_B T \gamma_T(t-t'), \\
			\mean{\eta (t)} = 0, \qquad \mean{\eta(t) \eta(t')}  = k_B T \gamma_R(t-t'), 
		\end{gather}
	\end{subequations}
	with the translational noise correlator $\gamma_T(t)$, and the rotational noise correlator $\gamma_R(t)$.
	Here, $\otimes$ is the dyadic product, $\id$ the identity matrix, $k_BT$ the thermal energy, and $\mean{\dots}$ denotes the noise average. 
	
	In discussing Eqs.~\eqref{eq:langevin_r} and \eqref{eq:langevin_phi}, we first suppose we are at zero temperature $T=0$ (no noise). 
	In this case, the velocity is identical to the active propulsion and the particle performs either linear or circular swimming motion. 
	Now we introduce fluctuations/noise in the system that kick the particle out of that particular situation. 
	Then there are two effects: first temporally correlated noise which perturbs the swimming motion and second dissipation incorporated in the memory kernels which lead to a relaxation back to the steady state. 
	
	For reasons of generality, we first do not imply any relation between the dissipation and the fluctuations in the system. 
	However, in the case of internal noise, the memory kernels are related to the correlation function of the noise via the second fluctuation-dissipation theorem, i.e., $\Gamma_T(t)=\gamma_T(t)$,  $\Gamma_R(t)=\gamma_R(t)$ \cite{KuboTH1985}.
	On the other hand, when fluctuation and dissipation come from different sources, the memory kernel and the noise correlator are independent \cite{WangT1999,DespositoV2008}. 
	This was explicitly realized in a recent experiment on magnetic active dumbbells where the rotational diffusivity was artificially enhanced with magnetic fields and therefore decoupled from the thermal bath \cite{SprengerFRAIWL2020}.
	
	The memory kernels $\Gamma_T(t)$ and $\Gamma_R(t)$ describe the viscoelastic response of the fluid and can be determined experimentally.
	Probably most commonly used are micro-rheological measurements on passive probe particles to extract the functional form of the memory kernel by tracking the particles mean-square displacement \cite{vZantenR2000,vdGuchtBKBCS2003}. 
	Alternatively, the memory kernel can be approximately linked to the shear relaxation modulus of the medium which can be measured with oscillatory shear experiments \cite{MasonW1995}.  
	
	Further, we point out that the stochastic process given by Eqs.~\eqref{eq:langevin_r} and \eqref{eq:langevin_phi} is defined stationary by setting the lower limit of the integral equal to $- \infty$ 
	(see Ref.~\cite{IndeiSCP2010} for a detailed discussion on the choice of the lower limit in the memory term).
	
	In Eq.~\eqref{eq:langevin_r}, the effective self-propulsion force is of the form $ \vec{F}_v(t) = v_0 \integral{-\infty}{t}{ \Gamma_T(t-t')  \uvec{n}(t') }{ t' } $.
	This choice is not unique but could in principle vary for different systems (for instance, externally actuated or mesoscopic swimmers).   
	In our model, we describe the force-free propulsion of a colloidal microswimmer which sets the fluid around itself in motion and translates in the resulting flow field.
	As a consequence, the propulsion force is linked to the viscoelastic response of the fluid and the internal active force $ \vec{F}_v(t)$ lags generally behind the orientation $\uvec{n}(t)$ \cite{NarinderBGS2018}. 
	
	Importantly, we remark that Eqs.~\eqref{eq:langevin_r} and \eqref{eq:langevin_phi} mark a special case of the model recently proposed by Narinder and coworkers \cite{NarinderBGS2018} which contains an additional torque proportional to the swim force, $\propto \uvec{n}(t) \times \vec{F}_v(t)$, explaining an increase of rotational diffusion \cite{GomezSBB2016} and the onset of circular trajectories \cite{NarinderBGS2018} for self-propelled Janus particles in a viscoelastic fluid. 
	Here we decouple the swim torque from the swim force at the benefit that we can solve the stochastic Langevin equations analytically.
	
	Last, the special case of active Brownian motion \cite{HowseJRGVG2007,vanTeeffelen2008,tHagenvTL2011} is recovered for instantaneous friction and zero-mean Gaussian white noise 
\begin{subequations} \label{eq:abp_limit}
    	\begin{gather}
    		\Gamma_T (t) = \gamma_T (t) = 2 \gamma_t \DiracDelta{t}, \\
    		\Gamma_R (t) = \gamma_R (t) = 2 \gamma_r \DiracDelta{t}, 
    	\end{gather}
\end{subequations}	
	where $\gamma_t$ and $\gamma_r$ are translational and rotational friction coefficient, respectively.
	 
\section{General results} \label{sec:General results}
	
    In this section we present analytic results for arbitrary memory kernel and noise correlator. 
    By calculating the Fourier transform of Eqs.~\eqref{eq:langevin_r} and \eqref{eq:langevin_phi}, a solution for the position $\vec{r}(t)$ and the orientation angle $\phi(t)$ can be derived as
	\begin{subequations}
		\begin{align}
			\vec{r}(t)  = & \, \vec{r}(t_0) + v_0 \integral{t_0}{t}{\uvec{n}(t')}{t'} \nonumber  \\ 
			& + \integral{-\infty}{\infty}{\big(\chi_T (t-t') -\chi_T (t_0-t') \big) \vec{\xi}(t')}{t'}, \label{eq:langevin_solution_r} \\
			\phi(t)  = & \, \phi(t_0) + \omega_0 (t-t_0) \nonumber   \\
			& + \integral{-\infty}{\infty}{\big(\chi_R (t-t') -\chi_R (t_0-t') \big) \eta (t')}{t'},   \label{eq:langevin_solution_phi}
		\end{align}
	\end{subequations}
	with the inverse Fourier transform of 
	\begin{subequations}
		\begin{gather}
			\tilde{\chi}_T(\omega) = \big(i \omega \tilde{\Gamma}_T^+(\omega) \big)^{-1}, \quad  \Gamma_T^+(t) = \Gamma_T(t)\Theta(t), \\
			\tilde{\chi}_R(\omega) = \big(i \omega \tilde{\Gamma}_R^+(\omega) \big)^{-1}, \quad  \Gamma_R^+(t) = \Gamma_R(t)\Theta(t),	
		\end{gather}
	\end{subequations}
	where we used the convention $\tilde{f}(\omega) = \integral{-\infty}{\infty}{f(t) e^{-\ii\omega t}}{t}$ for the Fourier transform of a function $f(t)$, and multiplied with the Heaviside function $f(t) \Theta(t)$, $\tilde{f}^+(\omega)$ yields the one-sided Fourier transform $\integral{0}{\infty}{f(t) e^{-\ii\omega t}}{t}$.

	The deterministic solution of Eqs.~\eqref{eq:langevin} (at zero temperature, $T=0$) is independent of the specific form of the memory kernel and the particle moves on either linear or circular trajectories
\begin{equation} \label{}
	\vec{r}(t) =
	\begin{cases}
		\vec{r}(0) + v_0 t \, \uvec{n}(0) , & \omega_0 = 0, \\
		\vec{r}(0) + \frac{v_0}{\omega_0} \big( \uvec{n}_\perp (0) - \uvec{n}_\perp (t) \big),  & \omega_0 \neq 0,
	\end{cases} 
\end{equation}	
	with $\uvec{n}_\perp (t) = \big(- \sin (\phi(0) + \omega_0 t),  \cos (\phi(0) + \omega_0 t) \big)^{T}$. 
    
    In the presence of noise, the motion of the particle can be characterized in terms of the low-order moments of the stochastic process.  
    Although Eq.~\eqref{eq:langevin_phi} is non local in time (and thus non-Markovian), the transitional probability for an angular displacements $\Delta \phi$ after a time $t$ is still Gaussian and specified by the mean $\mu(t) = \mean{\Delta \phi (t)}$ and the variance of the angular displacement $\sigma(t) = \mean{\Delta \phi^2 (t)} - \mean{\Delta \phi (t)}^2$ which are given by 
\begin{gather}
	\mu(t) = \omega_0 t, \\
	\sigma(t)  = \frac{k_B T}{\pi} \integral{-\infty}{\infty}{ \big(1 - e^{i\omega t}\big) \tilde{\gamma}_R(\omega) \tilde{\chi}_R(\omega) \tilde{\chi}_R(-\omega) }{\omega}. 
\end{gather}
    
    From that the orientation correlation function $C(t) = \mean{\uvec{n}(t) \cdot \uvec{n}(0)}$ can be readily derived and follows from
\begin{equation} \label{eq:orientational_correlation_function}
	\mean{\uvec{n}(t_2) \cdot \uvec{n}(t_1)} = \cos \big( \mu(\abs{t_2 - t_1}) \big) e^{ - \frac{1}{2}  \sigma ( \abs{t_2 -t_1})}.
\end{equation}
	Due to the stationarity of the underlying stochastic process, the two-time orientational correlation function only depends on the time difference.

	The general result for the mean displacement $\mean{\Delta \vec{r} (t)} = \mean{\vec{r}(t)- \vec{r}(0)}$ is
\begin{align}
	\mean{\Delta \vec{r} (t)} = v_0 \integral{0}{t}{ 	\mean{\uvec{n}(t') \vert \uvec{n} (0)} }{t'} , \label{eq:MD}
\end{align}
	where the conditional average
\begin{equation}
	\mean{\uvec{n}(t_2) \vert \uvec{n} (t_1)} = \uvec{P} \big[ e^{-\frac{1}{2} \sigma(t_2-t_1) + \ii ( \phi(t_1) + \mu(t_2-t_1)  ) } \big]  \label{} 
\end{equation}
	denotes the mean orientation at time $t_2$ under the condition that the particle had the angle $\phi(t_1)$ at previous time $t_1$ and $\hat{\vec{P}}[z] = \big(\text{Re}(z), \text{Im}(z)\big)^T$ transforms a complex number $z$ into its two-dimensional vector. 
	We remark, that the mean displacement is in general independent on the specific choice of the translational memory kernel $\Gamma_T(t)$ and only involves the coupling to the rotational dynamics of the particle.
	
	Next, the mean-square displacement is given by 
\begin{align}
	\mean{\Delta \vec{r} ^2 & (t)} =  v_0^2 \integral{0}{t}{\integral{0}{t}{ \mean{\uvec{n}(t') \cdot \uvec{n}(t'')}}{t''}}{t'} \label{eq:MSD} \\
	& +\frac{2 k_B T}{\pi} \integral{-\infty}{\infty}{ \big(1 - e^{i\omega t}\big) \tilde{\gamma}_T(\omega) \tilde{\chi}_T(\omega) \tilde{\chi}_T(-\omega) }{\omega}. \nonumber
\end{align}
	The first term describes the active contribution to mean-square displacement while the second term contains information on the passive translation caused by the noise (via $\gamma_T(t)$) and influenced by dissipation (via $\Gamma_T(t)$). 
	
	The effective self-propulsion force $\vec{F}_v(t)$ does not follow instantaneously the orientation of the particle. 
    It rather contains integrated information of past orientations and therefore lags behind $\uvec{n}(t)$. 
    To quantify the delay between the effective self-propulsion force and the particle orientation, we define the memory delay function 
\begin{equation}
    d(t) =  \mean{\vec{F}_v(t) \cdot \uvec{n}(0)} - \mean{\vec{F}_v(0) \cdot \uvec{n}(t)} \label{eq:delay}
\end{equation}     
    as the average difference between the projection of the active force $\vec{F}_v(t)$ on the initial orientation $\uvec{n}(0)$ and the projection of the orientation $\uvec{n}(t)$ and the initial active force $\vec{F}_v(0)$. 
	In Newtonian fluids, the effective self-propulsion force is proportional and instantaneous in the orientation, and thus the delay function equates to zero for all time.
	In a similar manner, the inertial delay function was previously defined for macroscopic active particles which measured the mismatch between the particle velocity $\dot{\vec{r}}(t)$ and the particle orientation $\uvec{n}(t)$ \cite{ScholzJLL2018,Loewen2020,SprengerJIL2021}.
	In our overdamped system, this inertial delay function is always zero since the average velocity is aligned with the orientation.
	Vice versa, for inertial particles subject to instantaneous friction the memory delay function vanishes.

	
	In the following section, we explicitly evaluate the introduced quantities for an exponential memory kernel and discuss the effect of memory on the dynamics of active Brownian particles. 
	
\section{Maxwell kernel}
	
	Arguably, the most prominently used memory kernel is given by the generalized Maxwell’s model (also know as Jeffrey’s  model) which adds additional exponential memory to the instantaneous friction \cite{PaulRB2018}. 
    For simplicity, we assume internal noise such that the memory kernels are related to the correlation functions of the noise via the second fluctuation-dissipation theorem. 
    Further, the same temporal dependency is adopted for the translation and the rotation, respectively,
    \begin{subequations} \label{eq:memory_exponential}
    	\begin{gather}
    		\Gamma_T (t) = \gamma_T (t) = \gamma_t \,\left( 2 \DiracDelta{t} + \frac{\Delta}{\tau} e^{-\abs{t}/\tau} \right), \label{eq:memory_exponential_trans} \\
    		\Gamma_R (t) = \gamma_R (t) = \gamma_r \, \left( 2 \DiracDelta{t} + \frac{\Delta}{\tau} e^{-\abs{t}/\tau} \right).	\label{eq:memory_exponential_rot}
    	\end{gather}
    \end{subequations}
    Here $\gamma_t$ and $\gamma_r$ denote reference translational and rotational friction coefficients, respectively. 
    The first term in Eqs.~\eqref{eq:memory_exponential_trans} and \eqref{eq:memory_exponential_rot} accounts for the instantaneous relaxation whereas the second term introduces the time-delayed response of the viscoelastic fluid with the relaxation time $\tau$ and the memory strength $\Delta$. 
    We remark, that for $\Delta=0$, $\tau \to 0$ or $\tau \to \infty$ the translation and rotational memory kernel get solely instantaneous and we recover the Markovian (no memory) active Brownian particle model \cite{HowseJRGVG2007,vanTeeffelen2008,tHagenvTL2011}. 
    
    Numerous rheological measurements have shown this Maxwell-like behavior in fluids including polymer solutions \cite{AnnableBEW1993,SprakelvdGCSB2008}, micelles \cite{CardinauxCSS2002,GalvanMiyoshiDC2008}, and cytoplasm \cite{WilhelmGB2003,Berret2016}.
    From the theoretical side, there exist several works which considered the effects of exponential memory on the Brownian motion of passive \cite{GrimmJF2011,RaikherRP2013} and active colloids \cite{GhoshLMM2015,NarinderBGS2018,SevillaRGS2019,MitterwallnerLN2020}.

\subsection{Orientation correlation function}
	
\begin{figure}
	\includegraphics[width=\columnwidth]{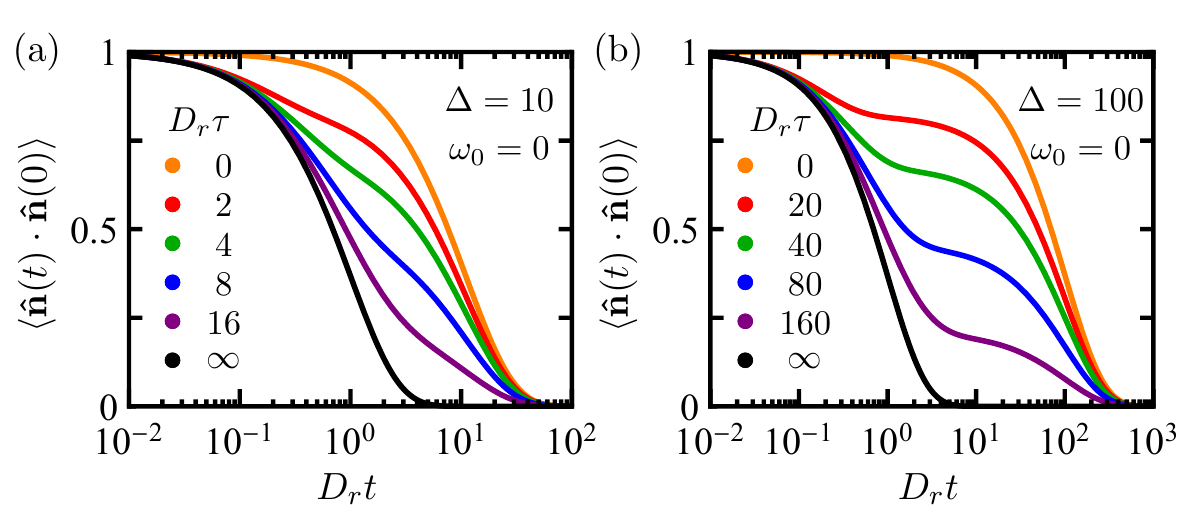}
	\caption{Orientation correlation $\mean{\uvec{n}(t) \cdot \uvec{n}(0)}$ as a function of $D_r t$ for different reduced relaxation times $D_r \tau$. Panel (a) is obtained with $\Delta = 10 $, and panel (b) with $\Delta = 100$. For $D_r \tau \to 0$ and $D_r \tau \to \infty$, the orientation decorrelates single exponentially. For in-between values, we find partial decorrelations at separated time-scales.}
	\label{fig:orientation_correlation}
\end{figure}	
	
The dynamical orientation correlation function $C(t)  = \mean{\uvec{n}(t) \cdot \uvec{n}(0)}$ has a double exponential structure 
\begin{equation}
	C(t)  = \cos(\omega_0 t) \, e^{- \frac{D_r}{1+\Delta}  \big( t +  \frac{\tau \Delta}{1+\Delta}  \big( 1 - e^{-(1+\Delta) t / \tau} \big) \big) },  \label{eq:orientation_correlation}
\end{equation}
with the short-time rotational diffision coefficient $D_r = k_B T / \gamma_r$. 
Eq.~\eqref{eq:orientation_correlation} simplifies to a single exponential decay for either small relaxation times $\tau$ or large ones
\begin{equation} \label{eq:orientation_correlation_asym}
	C(t)   \sim
	\begin{cases}
		\cos(\omega_0 t) \, e^{- D_r t} , & D_r \tau \gg 1+\Delta, \\
		\cos(\omega_0 t) \, e^{- \frac{D_r}{(1+\Delta)} t},  & D_r \tau \ll 1 + \Delta .
	\end{cases} 
\end{equation}
These Markovian (no memory) extreme cases are shown in orange $(\tau  \to 0)$ and black $(\tau \to \infty)$ in Fig.~\ref{fig:orientation_correlation}, where we plotted the orientation correlation for sufficiently large memory strength $\Delta$ and various values of $D_r \tau$.
We note that memory effects only occur when 
\begin{equation} 
	D_r \tau \simeq  (1+\Delta), \label{eq:memory_condition} 
 \end{equation}
in this case, we first see a partial decorrelation at time $1/D_r$, and a final decorrelation at a later time $(1+\Delta)/D_r$ (see Fig.~~\ref{fig:orientation_correlation}). 

A double-exponential for the orientation correlation was previously reported by Ghosch and coworkers \cite{GhoshLMM2015} and for inertial active particles \cite{WalshWSOBM2017,ScholzJLL2018}. 
Compared to these systems, we find different behavior for short times where the exponent is linear in time
\begin{equation}
    C(t)  =\cos(\omega_0 t) \, e^{- {D_r} ( t - \frac{\Delta}{2 \tau} t^2 + \mathcal{O} (t^3) ) }.
\end{equation}

One characterizing quantity of active particles is the persistence time $\tau_p = \integral{0}{\infty}{C(t)}{t}$ which denotes the average time the particle holds its orientation. 
Here the persistence time is evaluated as
\begin{equation}
	\tau_p  = \frac{\tau}{1+\Delta} \Re \big[  S^{-\Omega} e^{S} \Gamma (\Omega,0,S) \big] , \label{eq:persistence_time}
\end{equation}
with
\begin{equation}
	S  = \frac{ - \Delta \tau D_r }{(1+ \Delta)^2}, \quad \Omega = \frac{\tau}{1+ \Delta} \left( \frac{D_r}{1+\Delta} - \ii \, \omega_0 \right),
\end{equation}
and the incomplete gamma function $\Gamma (x,z_0,z_1) = \integral{z_0}{z_1}{t^{x-1} e^{-t}}{t}$.
Obvious from Eq.~\eqref{eq:orientation_correlation_asym}, the persistence time simplifies for small or large relaxation times $\tau$ to
\begin{equation} 
	\tau_p  \sim
	\begin{cases}
		\frac{D_r}{D_r^2+\omega_0^2} , & D_r \tau \gg 1+\Delta, \\
		\frac{D_r(1+\Delta) }{D_r^2+\omega_0^2(1+\Delta)^2},  & D_r \tau \ll 1 + \Delta,
	\end{cases} 
\end{equation}
representing the known result for active Brownian particle in simple Newtonian fluids \cite{vanTeeffelen2008,KurzthalerF2017,tHagenvTL2011}.

\subsection{Mean displacement}	

\begin{figure}
	\includegraphics[width=\columnwidth]{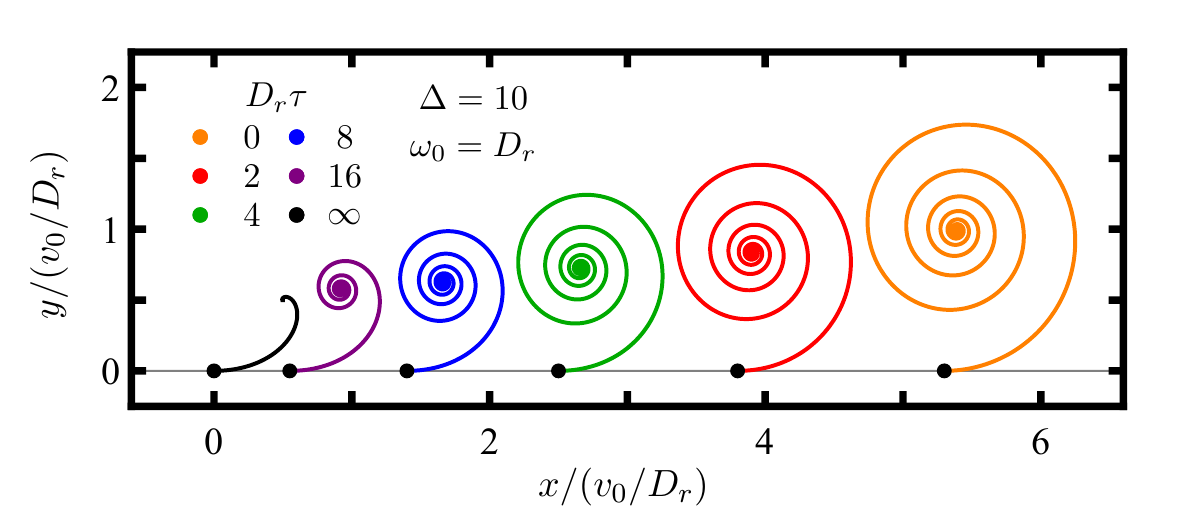}
	\caption{
		Mean displacement $\mean{\Delta \vec{r}(t)}$ in the $xy$-plane for $\Delta = 10$, $\omega_0=D_r$, and several values of $D_r \tau$.
		The initial orientation is set along the $x$-axis and the starting point at $t=0$ is denoted by a black dot.
		For $D_r \tau \to 0$ and $D_r \tau \to \infty$, the trajectory displays a perfect \textit{spira mirabilis}.
		}
	\label{fig:mean_displacement}
\end{figure}
	
Next, we address the mean displacement $\mean{\Delta \vec{r}(t)}$ for a given initial orientation $\phi(0)$ at $t=0$
\begin{equation}
	\mean{\Delta \vec{r}(t)} =   \frac{v_0\tau}{1+\Delta} \hat{\vec{P}} \Big[  S^{-\Omega} e^{S} \Gamma \big(\Omega,S e^{- (1+\Delta) t/ \tau},S\big) e^{i \phi(0)}\Big], 
\end{equation}
with the operator $\hat{\vec{P}}[z] = \big(\text{Re}(z), \text{Im}(z)\big)^T$. 
The mean displacement is increasing linearly for short times $\mean{\Delta \vec{r}(t)} =   v_0 t \uvec{n}(0) + \mathcal{O} (t^2)$ and saturates to finite persistence length 
\begin{equation}
	\lim\limits_{t\to\infty} \mean{\Delta \vec{r}(t)} =   \frac{v_0\tau}{1+\Delta} \hat{\vec{P}} \big[  S^{-\Omega} e^{S} \Gamma (\Omega,0,S) e^{i \phi_0}\big]. 
\end{equation}
We again mention that the mean trajectory is independent on the translational memory kernel noise  (see Eq.~\eqref{eq:memory_exponential_trans}) and only involves the coupling to the rotational dynamics of the particle (see Eq.~\eqref{eq:MD}).

In Fig.~\ref{fig:mean_displacement},  we show the mean trajectory of a circle swimmer ($\omega_0 \neq 0$) for different values of $D_r \tau$.
For very large relaxation times, the particle decorrelates before additional memory can prolong the persistence. 
Consequently, the mean trajectory displays a \textit{spira mirabilis} known for active particles in Newtonian fluids (see the black curve in Fig.~\ref{fig:mean_displacement}). 
When the relaxation time $\tau$ gets comparable to $(1+\Delta)/D_r$, the rotational friction gets enhanced at later times and circular motion gets more stable against noise perturbation (see the purple and blue curve in Fig.~\ref{fig:mean_displacement}).
Further decreasing relaxation time (see the green and red curve in Fig.~\ref{fig:mean_displacement}) the mean displacement approaches again the form of a \textit{spira mirabilis} with a decrease rotational diffusion coefficient $D_r/(1+\Delta)$ (see the orange curve in Fig.~\ref{fig:mean_displacement}).

\subsection{Mean-square displacement}	

	The mean-square displacement can be calculated as
	\begin{align}
		\mean{\Delta \vec{r} ^2 (t)} = & 4D_L t + \frac{4  \Delta \tau D_t }{(1+ \Delta)^2} \Big(1- e^{- (1+ \Delta)t/ \tau } \Big) \nonumber \\
		& - \frac{2  v_0^2 \tau^2 }{(1+ \Delta)^2} \Big( F(0) - F(t) \Big) ,  \label{eq:MSD_Maxwell}
	\end{align}
	with the long-time diffusion coefficient
	\begin{equation}
		D_L =  \frac{ D_t }{1+ \Delta}  + \frac{v_0^2 \, \tau}{2(1+ \Delta)} \Re \big[  S^{-\Omega} e^{S} \Gamma (\Omega,0,S) \big] ,  \label{eq:long_time_diffusion}
	\end{equation}
	and 
	\begin{align}
		F(t) = &  \text{Re} \Biggl\{ \frac{ e^{S } }{\Omega^2} \, \pFq{2}{2}{\Omega,\Omega}{\Omega+1,\Omega+1}{-S e^{- (1+\Delta) t/ \tau}} \nonumber \\
		& \times  e^{-(1+\Delta) \Omega t/ \tau }   \Biggr\}  , \label{}
	\end{align}	
	where $_qF_p$ represents the generalized hypergeometric function.
    In the passive case ($v_0=0$), the particle starts in a diffusive regime, $\mean{\Delta \vec{r} ^2 (t)} =  4 D_t t + \O{t^2}$, characterized by the short-time translational diffusion coefficient $D_t= k_B T /\gamma_t$ and then enters a sub-diffusive regime which leads to long-time diffusion with a reduced translational diffusivity $D_t/(1+\Delta)$. 
	Considering the active contribution ($D_t=0$) , the particle moves ballistic for short times, $\sim v_0^2 t^2$, and then undergoes a super-diffusive (or sub-ballistic) transition towards a long-time diffusive regime proportional to the speed square and the persistence time $ \sim v_0^2 \tau_p t / 2$.

\begin{figure}
	\includegraphics[width=\columnwidth]{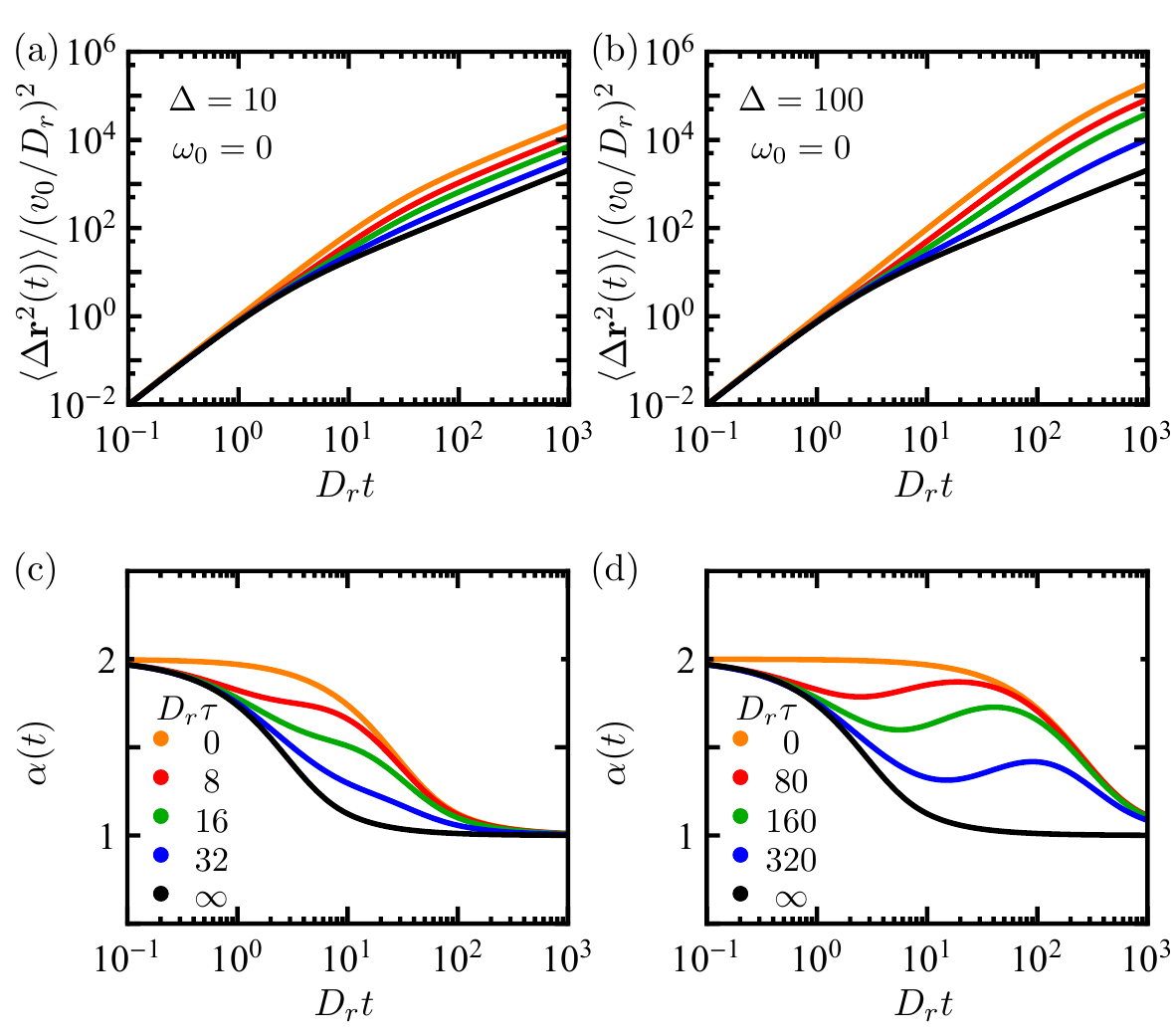}
	\caption{
		The mean-square displacement $\mean{\Delta \vec{r}^2(t) }$ and the corresponding dynamic exponent $\alpha(t)$ as a function of time $t$ for several values of $D_r \tau$. Panels (a) and (c) are obtained with $\Delta = 10$, and panels (b) and (d) with $\Delta = 100$.}
	\label{fig:mean_square_displacement}
\end{figure}
\begin{figure*}
	\includegraphics[width=1.666\columnwidth]{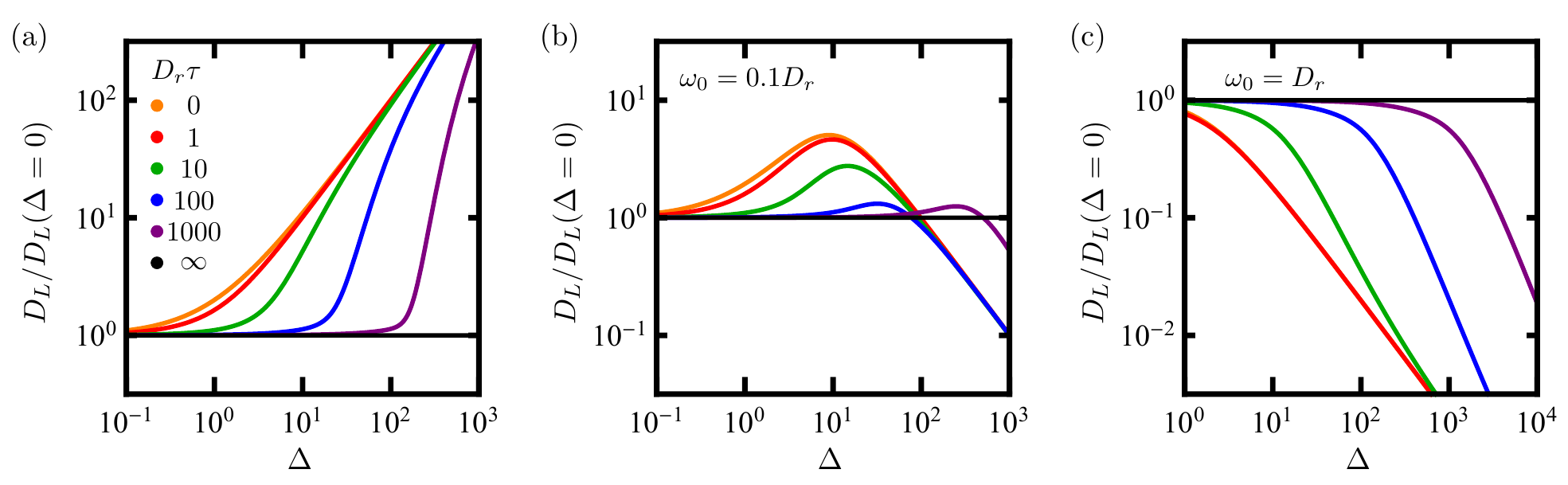}
	\caption{Long-time diffusion coefficient $D_L$ as a function of the memory strength $\Delta$ for several values of $D_r \tau$  and different circling frequencies $\omega_0=0$ (a), $\omega_0 = 0.1 D_r$ (b), and $\omega_0 = D_r$ (c).
		The translational diffusion coefficient was set to zero, $D_t=0$.}
	\label{fig:long_time_diffusion}
\end{figure*}
	
	In Fig.~\ref{fig:mean_square_displacement}, we plot the active contribution of the mean-square displacement ($D_t=0$) for two values of the memory strength $\Delta$ over the range of relevant values of $D_r \tau$, and also show the corresponding dynamic exponent given by the logarithmic derivative  
\begin{equation}
    \alpha(t) = \frac{\dif \log( \mean{\Delta \vec{r} ^2 (t)}  ) }{ \dif \log(t) }  .
\end{equation}	
	The dynamic exponent $\alpha(t)$ is able to resolve the relevant time scales of the system more clearly: if, for example, the mean-square displacement follows a power law $\mean{\Delta \vec{r} ^2 (t)} \sim t^\alpha$, $\alpha(t)$ is equal to the power-law exponent $\alpha$.
	For the Markovian extreme cases ($\tau \to 0$ and $\tau\to \infty$), we find a clean transition from a ballistic regime ($\alpha = 2$) to a diffusive one ($\alpha =1$).
	For in-between values of $D_r\tau$, the dynamic exponent $\alpha(t)$ starts decreasing when the first decorrelation happens at times $t \gtrsim 1/D_r$.
	If the memory strength $\Delta$ is sufficiently large (see  Fig.~\ref{fig:mean_square_displacement}(d)), the dynamic exponent is increasing again at times $t \gtrsim \tau /(1+\Delta)$. 
	This event coincides with the persistent plateau in the orientation correlation function (see  Fig.~\ref{fig:orientation_correlation}(d)). 
	Finally, the particle transitions to a diffusive regime ($\alpha = 1$) for times $t \gtrsim (1+\Delta)/D_r$.
	
	The long-time diffusion coefficient $D_L$ (see Eq.~\eqref{eq:long_time_diffusion}) depends non-trivially on the parameter of the model. 
	In Fig.~\ref{fig:long_time_diffusion}, we show the long-time diffusion coefficient as a function of the memory strength $\Delta$ and various values of $D_r \tau$. 
	For a vanishing circling frequency ($\omega_0 = 0$), the long-time diffusion coefficient is monotonically increasing as a function of the memory strength $\Delta$ and monotonically decreasing as a function of the relaxation time $\tau$ (see Fig.~\ref{fig:long_time_diffusion}(a)).
    However, for a finite relaxation time, the asymptotic behavior of the long-time diffusion coefficient for large $\Delta$ is given by $D_L \sim v_0^2 \Delta / (2 D_r)$.  
    For small circling frequency  (see Fig.~\ref{fig:long_time_diffusion}(b)), the long-time diffusion behaves non-monotonic in $\Delta$. 
    The optimal memory $\Delta_\text{opt}$ is increasing as a function of relaxation time $\tau$ while the corresponding maximal value $D_L(\Delta_\text{opt})$ is decreasing. 
    At larger circling frequency (see Fig.~\ref{fig:long_time_diffusion}(c)), the long-time diffusion decreases immediately as a function of $\Delta$, $D_L \sim v_0^2 D_r  / (2 \Delta \omega_0^2)$.

\subsection{Delay function}

\begin{figure}
	\includegraphics[width=\columnwidth]{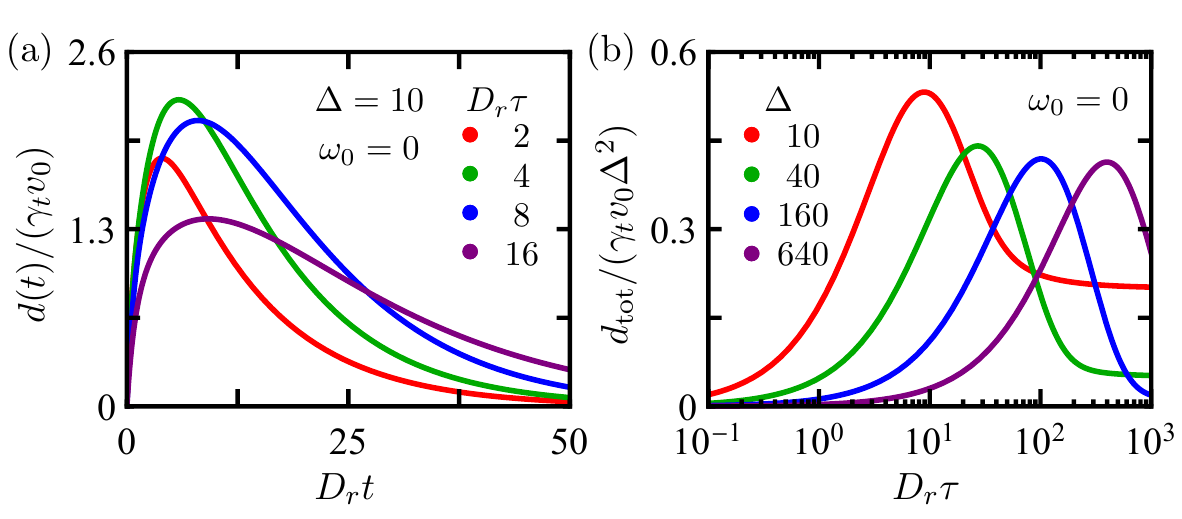}
	\caption{Panel (a), memory delay function $d(t)$ as a function of $D_r t$ for different reduced relaxation times $D_r \tau$ and  $\Delta = 10$. Panel (b), total delay $d_\text{tot}$ weighted with $\Delta^2$ as a function of the reduced relaxation time $D_r \tau$ for different values of the memory strength $\Delta$.}
	\label{fig:delay_function}
\end{figure}
    
    In Eq.~\eqref{eq:delay}, we defined the memory delay function $d(t)$ to quantify the memory induced mismatch between the effective self-propulsion force $\vec{F}_v(t)$ and the particle  orientation $\uvec{n}(t)$. Evaluated for the Maxwell kernel, we find
	{\allowdisplaybreaks
		\begin{align}\label{eq:delay_function}
			d(t)  = & \gamma_t v_0  \frac{\Delta e^{ S }}{1+\Delta}  \Re \bigg\{ S^{-\Omega_{+} } \Big( \Gamma\left(\Omega_{+},0, S\right) e^{-t/\tau} \nonumber \\
			& - \Gamma\left(\Omega_{+},0, S e^{-(1+\Delta)t/\tau}\right) e^{t/\tau} \Big) \nonumber \\
			&+ S^{-\Omega_{-} } \Gamma\left(\Omega_{-}, S e^{-(1+\Delta)t/\tau}, S \right) e^{-t/\tau} \bigg\}, 
	\end{align}}%
	with 
	\begin{equation}
		\Omega_{\pm} = \frac{\tau}{1+ \Delta} \left( \frac{D_r}{1+\Delta} \pm \frac{1}{\tau} - \ii \, \omega_0 \right),
	\end{equation}
	The memory delay function is constructed such that it vanishes when the translational memory function responses instantaneously (meaning, $\Gamma_T(t) = 2 \gamma_t \DiracDelta{t}$). 
	Thus, consistent with previous considerations, $d(t)$ vanishes for the Markovian limits of the model $\Delta = 0$, $\tau \to 0$, and  $\tau \to \infty$.
	In Fig.~\ref{fig:delay_function}(a), we show the delay function $d(t)$ as a function of time for in-between values of $D_r \tau$.
	The delay function is always positive for linear swimmer ($\omega_0 = 0$), starts at zero, has a positive peak $d(t_\text{opt})$ after a typical delay time $t_\text{opt}$, and decorrelates to zero for long times. 
	Both the peak value and the typical delay time depend non-monotonic on the relaxation time  $\tau$ and show a single maximum around $D_r \tau \simeq  (1+\Delta)$ (recalling the condition for memory effects Eq.~\eqref{eq:memory_condition}). 
	
	We define the total delay of the particle as $d_\text{tot} = \integral{0}{\infty}{d(t)}{t}$ which yields 
\begin{equation}\label{eq:total_delay}
		d_\text{tot}  =  \gamma_t v_0 \tau \frac{2 \Delta e^{ S }}{1+\Delta}  \Re \big[ S^{-\Omega_{+} } \Gamma\left(\Omega_{+},0, S\right) \big], 
\end{equation}
    and is shown in Fig.~\ref{fig:delay_function}(b) as a function of the reduced relaxation time $D_r \tau$.  
    Similar to the peak value $d(t_\text{opt})$, the total delay becomes maximal around $D_r \tau \simeq  (1+\Delta)$.
    For representative reasons, we decided to weight the total memory by the memory strength square, i.e, $d_\text{tot}/(\gamma_t v_0 \Delta^2)$ in Fig.~\ref{fig:delay_function}(b). 
    In that way, we find that $d_\text{tot} \sim \Delta^2$ around the relevant values of $D_r \tau$ (see Eq.~\eqref{eq:memory_condition}).
    Although $d(t) \to 0$ for $\tau \to \infty$, the total memory saturates to the non-zero value $d_\text{tot} \sim 2 \Delta \gamma_t v_0$ for $\tau \to \infty$ (limit and integral do not commute in this case).

	\section{Conclusions}
	
	In this work, we studied a self-propelled colloid in a viscoelastic medium. The particle itself was modeled in terms of non-Markovian Langevin equations which included memory effects in the particle friction to account for the viscoelastic background.
	Analytical solutions are presented. 
	This model may serve as a benchmark and simple framework to evaluate and interpret experimental or simulation data for particle trajectories obtained in realistic and more complex environments \cite{SaadN2019}.
	In particular the nature of the memory kernel can in principle be determined by fitting the experimental correlations to the solutions of our model corresponding to micro-rheology \cite{MasonGKFW2000,GazuzPVF2009,HeidenreichHK2011,PuertasV2014,MalgarettiPP2022}. 
	
	We evaluated our general results explicitly for the Maxwell kernel which adds exponentially decaying memory to the standard instantaneous Stokes friction.
	In particular, we found a double exponential structure for the orientational correlation function exhibiting partial decorrelation at short times and the existence of persistent plateaus for intermediate times.
	In order for memory effects to occur, we identified a relation between the short-time rotational diffusion coefficient, the memory strength and corresponding relaxation time (see Eq.~\eqref{eq:memory_condition}) and discussed the influence of memory at intermediate and long timescales for the mean and mean-square displacement of the particle. 
	Last, we quantified the delay between effective self-propulsion force and the particle orientation in terms of a new defined memory delay function.
	
	Our model can be extended to higher spatial dimensions \cite{tHagenvTL2011}, to harmonic confinement \cite{DespositoV2009,Szamel2014,JahanshahiLtH2017,CapriniSLW2022}, to external fields \cite{tenHagenKWTLB2014,AbdoliVWSBS2020}, and to include inertia \cite{ScholzJLL2018,Loewen2019, Sandoval2020, BreoniSL2020, Loewen2020, CapriniMBM2021, SprengerJIL2021} where an analytical solution seems to be in reach as well. 
	Moreover different combinations of friction and memory kernel as well as colored noise can be considered for future work \cite{MetzlerK2000,ZaidDY2011,HoflingF2013,GernertLLK2016,LoosHK2019}, for instance, Mittag-Leffler noise \cite{VinalesD2007,FigueiredoCamargoCdOV2009} or power-law memory \cite{MinLCKX2005,GomezSS2020}.
	Finally the collective behavior of many interacting active particles in a viscoelastic medium \cite{BozorgiU2013,LiebchenL2017,PaliwalRvRD2018, MurchS2020,teVrugtJW2021,HolubecGLKC2021,MaN2022} needs to be explored more and will be an important area of future research.
	
	\section{Acknowledgments}
	
	This work was supported by the SPP 2265 within the project LO 418/25-1.

	\bibliographystyle{apsrev4-1}
	\bibliography{refs}
	
\end{document}